# Field emission: applying the "magic emitter" validity test to a recent paper, and related research-literature integrity issues

Running title: Magic emitter validity test
Running Authors: Forbes

Richard G. Forbes[a)]

University of Surrey, Advanced Technology Institute & School of Computer Science and Electronic Engineering, Guildford, Surrey GU2 7XH, UK

[a)] Electronic mail: r.forbes@trinity.cantab.net

This work concerns studies of field electron emission (FE) from large area emitters. It discusses—and where possible corrects—several literature weaknesses related to the analysis of experimental current-voltage data and related emitter characterization, using a recent paper in Applied Surface Science to exemplify these weaknesses. One weakness, not detected in the published paper, is that current-density experiments and related theoretical predictions there differ by a large factor, in this case of order $10^{16}$. The work also shows that a recently introduced validity test—the "magic emitter" test—can be used, at the immediate-pre-submission or review stages, to help uncover scientific problems. More generally, in the literature of FE from large area emitters over the last 15 years or so, there appear to be many papers (perhaps hundreds of papers) with some or all of the weaknesses discussed. The scientific integrity of this research area, and of the related peer review processes, appear to be significantly broken, and attempts to correct the situation by the normal processes of science have had limited effect. There seems a growing case for independent wider investigation into research integrity issues of this general kind, and possibly for later action by Governments.

## I.  INTRODUCTION: BACKGROUND AND MOTIVATION

Large area electron emitters based on field electron emission (FE) have many actual and potential technological applications. In related materials development activities, it is important that the current-voltage characteristics of such emitters be analyzed correctly and that reliable values of emitter characterization parameters (particularly dimensionless field enhancement factors) be reported. A specific aim of the present paper is to draw attention to apparent weaknesses in data analysis in a recent paper[1] and in the review process that has not discovered these weaknesses. The paper under discussion is a particularly clear example of these weaknesses, which appear to have been widespread in sectors of field electron emission (FE) technology literature for the last fifteen years or more.  It will be shown that one serious weakness of this kind can be rapidly and easily uncovered, in the immediate-pre-submission or review stages of a paper, by applying, to reported experimental and characterization data, a newly developed validity test[2] provisionally called the "magic emitter test". The present discussion thus provides an illustration of the usage of this test. Related weaknesses, also found in Ref. 1, are indicated and discussed.

A much fuller discussion of one of the main underlying issues involved (whether an FE system is "electronically ideal") can be found in a recent paper[2], and a discussion of some less-critical issues appears in an earlier paper[3].

Most FE technology papers (including the paper under discussion and others recently published) analyze FE current-voltage [$I(V)$] experimental data using: (a) a methodology—the Fowler-Nordheim (FN) plot—originally developed[4] in 1929; and (b) a modern simplified version of the FE equation that was originally developed by Fowler and Nordheim[5] in 1928 and clarified by Stern et al.[4] in 1929. This simplified version omits a pre-factor of order unity that appears in the original 1928/29 equation, and is usually an acceptable mathematical approximation of 1928/29 FE theory.

Underlying this widely used methodology is the unstated assumption that the analysis of the data is a problem involving only the emission physics and the (zero-current) system electrostatics. An emitter for which this assumption is adequately valid has been termed *electronically ideal* (see Ref. 2). The metal emitters under consideration in 1929 could usually be treated as electronically ideal, but this is not necessarily true for modern emitters fabricated from non-metallic materials. In such cases it may often be necessary to take into account wider aspects of the electrical engineering of the complete emission and measurement system (see Ref. 2). Twelve different possible causes that can make FE systems *not* electronically ideal have been identified in Ref. 2. In any particular non-ideal case, more than one of these might be operating. If an FE system is not electronically ideal, then it is not scientifically valid to apply 1929 methodology: hence, results derived by using it inappropriately—in particular the values of dimensionless field enhancement factors (FEFs)—may be spurious.

As already indicated, large area field electron sources have some interesting actual and potential technological applications. Field enhancement factors are one of the parameters used to characterize newly developed emitting materials for such sources, and it is important for research-and-development purposes to have accurate FEF-values reported in the literature. But it is the author's belief that many FEF-values published in FE technological literature are in fact spurious. A small survey[6], conducted about ten years ago, provisionally suggested that up to about 40% of published FEF-values might be spurious.

## II. APPLICATION OF THE "MAGIC EMITTER" TEST

In order to deal with this situation, the present author has introduced the idea of a *validity check* (also called a *validity test*). This is an engineering type test that analyses the measured current-voltage data in order to establish whether the measured data show that the system is not (or may not be) electronically ideal. One of the original validity checks was the so-called *orthodoxy test*[6]. Recently[2] I have proposed a new validity check, the "magic emitter" test. This can be seen as a simplified version of the orthodoxy test that can be rapidly applied to published data when information is provided both about the range of macroscopic-field magnitudes ($F_M$) apparently used in the experiments and about the derived value of the apparent characteristic FEF. (This is denoted here by $(\gamma_{MC})^{app}$, but often by $\beta$ in the literature.)

When this information is provided, then the largest reported value $F_M(\text{max})$ of macroscopic-field magnitude can be used to derive the equivalent value $F_C^{\text{app}}(\text{max})$ of the apparent characteristic local surface-field magnitude, via the equation

$$F_C^{\text{app}}(\text{max}) = (\gamma_{MC})^{\text{app}} \cdot F_M(\text{max}) . \qquad (1)$$

For example, in Ref. 1, for their sample S3, the lowest reported value (see Fig. 6) of $1/F_M$ is about 0.17 µm/V, which means that $F_M(\text{max})$ is about 5.9 V/µm. The reported value of $(\gamma_{MC})^{\text{app}}$ is 14594, which yields the value $F_C^{\text{app}}(\text{max}) = 8.6 \times 10^4$ V/µm = 86 V/nm.

It is well established that if the field applied to a field electron emitter is gradually increased in magnitude then eventually the emitter will self-destruct, though precisely how this happens may depend on particular circumstances. A parameter of interest is the "*reference* local field-magnitude" $F_R(\phi)$ at which—for an emitter of characteristic local work function $\phi$—the top of a Schottky-Nordheim tunnelling barrier is pulled down to the emitter Fermi level. Sample S3 is stated to have a true work function of 3.25 eV; the corresponding reference-field magnitude is about[7] 7.3 V/nm. Usual thinking is that when the barrier top is at the Fermi level then the electrons certainly pour out of the emitter in such numbers that the resulting current density creates heating effects that rapidly destroy the emitter. However, older investigations[8] of FE in a traditional field electron microscope configuration suggest that (in that context) the highest safe continuous-current situation corresponds to a field magnitude significantly lower than $F_R(\phi)$, namely 0.34 $F_R(\phi)$ (see Ref. 7]). For a $\phi = 3.25$ eV emitter, this is a local surface-field magnitude of about 2.5 V/nm.

The original logic behind the "magic emitter" test was that if the value of $F_C^{app}(max)$ found from eq. (1) is significantly greater than the range of surface-field-magnitude values where the emitter is expected to self-destruct, as it is for sample S3, then authors apparently have a "magic emitter" immune to normal self-destruction processes. In reality, of course, authors in this position have failed to carry out a validity check, have applied 1929 data-analysis methodology in circumstances where it is not scientifically valid to do so, and are reporting spurious (unreasonably high) FEF values.

My revised thinking is that (for consistency) it is better to link the decision criterion in the magic-emitter test directly to the related criterion used in the orthodoxy text. This states that the extracted FEF-value is almost certainly spurious if the derived value of $F_C^{app}(max)$ is greater than $f_{ub}(\phi) \cdot F_R(\phi)$, where $f_{ub}(\phi)$ is a work-function-dependent parameter tabulated in Table 2 of Ref. 6. For $\phi=4.50$ eV we have $f_{ub}=0.75$ (which is the value usually quoted), but for $\phi=3.25$ eV the slightly higher value of $f_{ub}=0.88$ should be used. Sample S3 has a derived value of $F_C^{app}(max)$ equivalent to about 12 $F_R(\phi)$, so—as already concluded—the extracted FEF value is clearly almost certainly spurious.

When the derived characteristic local-field magnitude is close to $f_{ub}(\phi) \cdot F_R(\phi)$ (either above or below this value), then more careful investigation is needed.

## III. OTHER WEAKNESSES

### A. *The definition and evaluation of "current density"*

Two other weaknesses of the paper under discussion deserve comment, because (like the weakness discussed above) they seem to be widespread in FE technological literature. Equation (1) in Ref. 1 can be re-written in the direct form

$$J = A\phi^{-1}(\beta E)^2 \exp[-B\phi^{3/2}/\beta E] , \qquad (2)$$

where the symbols have the same interpretations as in Ref.1. If the values $A$=1.54 µA eV V$^{-2}$, $B$= 6.83×10$^7$ eV$^{-3/2}$ V cm$^{-1}$, $\phi$= 3.25 eV, $E$= 5.9×10$^4$ V/cm, $\beta$ = 14594 are substituted into eq. (2) here, then this yields a theoretical current-density prediction of around $J$ = 2×10$^{17}$ µA/cm$^2$. By contrast, Fig. 6(a) in Ref. 1 shows that the actual experimental result for sample S3, for $E$= 5.9 V/µm, is around 7 µA/cm$^2$. Thus, neither the authors nor the reviewing system has been able to detect an apparent discrepancy of order 10$^{16}$ between experiments described in Ref. 1 and theory stated in Ref. 1.

Part of this discrepancy is presumably a side-effect of the weakness discussed above, but there is certainly also another partial cause. Equation (2) above is in fact an equation for *local emission current density* (LECD). Because field enhancement is included, this equation is being applied in the high field region near the apex of a spike-like emitter. Thus, what eq. (2) actually predicts (using 1920s style theory) is a characteristic LECD $J_C$ at or near the emitter apex. By contrast, the symbol "$J$" on the vertical axis of Fig. 6(a) in Ref. 1 represents the macroscopic (or "footprint average") current density $J_M$. This $J_M$ is expected to be much lower than $J_C$. For electronically ideal emitters the ratio $J_M/J_C$ could be as low as 10$^{-7}$, or lower: however, actual values of this ratio are likely to be very variable as between different materials and different fabrication methodologies, and are not well known.

This failure to distinguish properly between local and macroscopic current densities appears to be widespread in FE technological literature, and the author has detected other papers that include large undiscussed apparent discrepancies between theory and experiment (though usually not so extreme as in the present case).

## *B.    Use of obsolete 1920s-style emission equations*

A further difficulty with eq. (2) above, and with the related equation in the paper under discussion, is that these equations represent versions of the FE theory that was developed in the late 1920s. In the 1950s serious errors were found[9] in all earlier FE theory, in that exchange-and-correlation effects (normally modelled as image-force effects) had not been taken into account in a physically and mathematically correct fashion. This resulted in a 1956 reformulation of FE theory by Murphy and Good[10] (MG). Some consequences in principle of this reformulation have been discussed elsewhere[3].

In the case under discussion, the main effect of this reformulated theory is to put into the exponent of eq. (2) a correction factor that (for a $\phi$ = 3.25 eV emitter) has values 0.8 down to 0.4, over the orthodoxy-test pass range of scaled field ($f$) between 0.175 and 0.525. This results in predictions of local current density that are larger, by a factor from around 1150 down to around 350, than those provided by simplified 1920s-style equations.

When both the above corrections are taken into account, then the equation for macroscopic current density $J_M$, in terms of the relevant local work function $\phi$ and the characteristic local surface-field magnitude $F_C$, can be written

$$J_M = \alpha\, t_F^{-2}\, a\phi^{-1}\, F_C^2\, \exp[-v_F\, b\phi^{3/2}/F_C]\,. \qquad (3)$$

Here: the first and second Fowler-Nordheim constants are represented by the lower-case letters $a$ and $b$, rather than by $A$ and $B$; $v_F$ and $t_F$ are appropriate particular values (appropriate to a Schottky-Nordheim (SN) tunnelling barrier defined by $\phi$ and $F_C$) of well-known special mathematical functions "v" and "t" (see any of Refs 10-13); and $\alpha$ is a parameter best called an *area efficiency*. This parameter $\alpha$ is a measure of how much of the footprint area of a large-area field electron source is actually emitting electrons. If required, the surface-field magnitude $F_C$ can be expressed in terms of a macroscopic-field magnitude by using an appropriate field enhancement factor.

The present author's view is that it is often better to replace this equation by one for the emission current $I_e$, which will be equal to the measured current if there is no significant leakage current. The replacement equation is

$$I_e = A_f^{SN}\, a\phi^{-1}\, F_C^2\, \exp[-v_F\, b\phi^{-1}/F_C]\,. \qquad (4)$$

where $A_f^{SN}$ is a parameter that the author calls the *formal emission area for the SN barrier*. For algebraic simplicity, the prefactor $t_F^{-2}$ and also temperature effects in MG FE theory[10] have been swept into $A_f^{SN}$. With electronically ideal FE systems, the parameter $A_f^{SN}$ can be extracted from experiments, but in the present state of theoretical knowledge it is not possible to make accurate predictions of its value for real emitters (see Ref. 14 for a more detailed discussion) nor to use it to make reliable estimates of actual emitting areas. Nonetheless, $A_f^{SN}$ can serve as a qualitative guide to the actual emitting area and could be a useful empirical characterization parameter.

Note that, because the area-value extracted in data analysis depends in principle on the barrier model assumed, it is necessary to choose a name for the extracted area that specifies explicitly what barrier has been assumed in the data analysis.

For comparison purposes, for electronically ideal FE systems, a more useful empirical parameter may be the *formal area efficiency for the SN barrier*, $\alpha_f^{SN}$, defined by

$$\alpha_f^{SN} \equiv A_f^{SN}/A_M , \qquad (5)$$

where $A_M$ is the *macroscopic area* (or "footprint") of the large-area emitter.

For clarity, it is strongly recommended that eq. (4) be called the *Murphy-Good FE equation* or (perhaps better—see Ref. 15) the *Extended Murphy-Good FE equation*. This would clearly distinguish eq. (4) from the 1928/29 Fowler-Nordheim FE equation. There is some tendency in the literature for theoreticians and some experimentalists to apply the name "Fowler-Nordheim equation" to the 1956 Murphy-Good FE equation. The present author finds this nomenclature convention very unhelpful, and suspects that it may be a partial cause of the literature confusion between the 1928/29 and 1956 FE equations (which, as just indicated, are significantly different in their numerical predictions).

## IV. COMMENTARY: WIDER IMPLICATIONS

It needs to be made clear that these comments should be seen, not as a specific criticism of the authors of the paper under discussion or of the reviewing system operated by Applied Surface Science, but rather as comments on the scientific state of FE technological literature as a whole, *as exemplified by* the paper under discussion. It is the present author's perception that similar weaknesses (though often less extreme numerically) can be found in many papers (perhaps hundreds of papers) and in the reviewing systems of many journals.

Sociological discussions of scientific-community behavior sometimes use the term "pathological" to describe situations where individual scientists or communities of scientists believe things and/or carry out procedures that scientists outside the community do not accept as valid. (The term "pathological" derives from the paper by Langmuir and Hall[16], but its meaning has become extended over time.)

A characteristic associated with pathological scientific communities is that members normally mainly cite work performed by other members of the community, and do not cite relevant work by scientists outside the community. The paper[1] under discussion in fact illustrates this: no citation is given to any modern paper on FE theory or to any of the 15 or so FE research handbooks/textbooks that have been published since 1960, all of which discuss Murphy-Good FE theory (for a list see Ref. 13).

This pathological literature situation has been ongoing in FE technological contexts for around 15 years. The research integrity of this research area is now significantly broken, and the peer review system in this research area is also significantly broken. Attempts by individual scientists over the last ten years to improve the situation by using the normal correction processes of science have had only limited effect.

The phenomenon of field electron emission is the basis for many established and potential technologies (some very credible although some very speculative). It is not helpful, and arguably it is not in the public interest, that apparently relevant published scientific literature contains large amounts of defective and misleading information. This situation is particularly unhelpful for non-experts, especially those interested in medical and defense applications, and especially new FE graduate students. The present author believes that, in individual countries such as the UK and USA, there is a growing case for some form of parliamentary, congressional or other official investigation, with a focus on how to safeguard the integrity of scientific literature published in that country. This should probably take place in the context of a wider investigation into research integrity and the difficulties of ensuring it. The field electron emission situation could perhaps be a useful example of possible wider problems of this general kind, because the issues in the FE case are very clearly defined.

Two final points need to be made. First, it is not being claimed here that Murphy-Good FE theory is "correct physics": clearly it is not, because it is a theory of FE from metals with smooth planar surfaces, and also disregards the effects of atomic structure. What is being claimed here is that MG FE theory is "obviously better physics" than 1920s style FE physics, and is "good enough" for data analysis in most modern technological contexts. When emitters are sharply curved then this may not be true, and the approach developed by Kyritsakis[17] may well be preferable.

Second, I emphasize again that this a *community problem* affecting the FE technology community interested in materials development for large-area field electron sources. It seems that hundreds of authors and reviewers, and very many journal editors, have been "hoaxed" by the related FE literature over the last 15 years or so. It would be inappropriate (and unfair) to attribute significant fault to the particular authors and journal used here to exemplify a very much wider problem. They, like many others, have been "hoaxed".

## V. SUMMARY AND CONCLUSIONS

In summary, the main points being made in this paper are the following.

– There are several significant weaknesses in the data analysis in the paper[1] under discussion, and (obviously) in the related peer review process. Where practicable, the present paper corrects these weaknesses.

– The "magic emitter" validity test can be a quick and easy method of uncovering the existence of a data-analysis problem, prior to submission or in the review process.

– It is the author's perception that similar weaknesses (and others) appear in many papers (perhaps in hundreds of papers in the last 15 years or so) in the FE technological literature relating to materials development for large area field electron sources. Research integrity and the peer review process in this specialist area both seem significantly broken.

– There seems a growing case for some sort of "official" external investigation into the situation described, as part of a wider investigation into research integrity in scientific literature and the difficulties of ensuring it.

# AUTHOR DECLARATIONS

**Conflicts of Interest**

The author has no conflicts to disclose.

# DATA AVAILABILITY

The data that support the findings of this study are available within the article.